\documentclass[twocolumn,journal]{IEEEtran}
\usepackage{graphicx}
\usepackage{amsmath}
\usepackage{amssymb}
\usepackage{tabu}
\usepackage{comment}
\usepackage{multicol}
\usepackage{amsthm}
\usepackage{cite}
\usepackage{float}
\usepackage{color}
\usepackage{balance}
\usepackage[outdir=./]{epstopdf}

\begin{document}
\title{ \vspace{-0.8cm} \LARGE Securing Full-Duplex Amplify-and-Forward Relay-Aided \\ Transmissions Through Processing-Time Optimization}

\author{Mohamed Marzban$^\dagger$, Ahmed El Shafie$^{\ddagger}$, Ahmed Sultan$^\star$, Naofal Al-Dhahir$^\dagger$
\begin{tabular}{c}
\small $^\dagger$University of Texas at Dallas, USA. \\
\small $^\ddagger$Qualcomm Inc, CA, USA. \\
\small $^\star$King Abdullah University of Science and Technology (KAUST), Saudi Arabia.
\end{tabular}
\vspace{-0.5cm}
}
\date{}

\maketitle


\begin{abstract}
We investigate physical-layer security of the full-duplex (FD) amplify-and-forward (AF) relay channel. We provide a new perspective on the problem and show that the processing time (delay) at the relay can be exploited to improve the system's security. We show that the FD AF relay channel can be seen as an intersymbol-interference (ISI) channel, hence, the discrete-Fourier transform (DFT) can be used for data modulation and demodulation to convert the frequency-selective channel into flat-fading channel per sub-channel/sub-carrier. By exploiting the fact that the channel memory needs to be cleared by inserting the cyclic-prefix, Alice injects an artificial-noise (AN) signal that hurts the eavesdropping nodes only. The strength of this AN signal and its interference rank are controlled by the relay's processing time.
\end{abstract}

\vspace{-0.3cm}
\section{Introduction}
Full-duplex (FD) amplify-and-forward (AF) relaying is a low-complexity approach utilized for enhancing wireless communication networks coverage and data rates \cite{8719975,ali2019full}. Relay-aided physical-layer (PHY) security has been investigated in many previous works \cite{huang2015buffer,7296606,chen2015physical,shafie_FD_HDhybrid, chen2019secure, cao2019secrecy}. In \cite{huang2015buffer}, Huang \emph{et al.} studied the PHY security of a half-duplex (HD) relay channel equipped with a data buffer. Considering a single relay, a single source node (Alice) and a single intended destination (Bob), an adaptive link selection scheme for time slot assignments between Alice and the relay was proposed in \cite{huang2015buffer} to protect the data against eavesdropping. The authors of \cite{7296606} investigated two cooperative secure schemes for a wireless relay channel with buffered relay and source nodes. 

 Recently, \cite{chen2015physical} investigated the PHY security of a single-input single-output single-antenna eavesdropper wiretap channel. 
The work of \cite{shafie_FD_HDhybrid} investigated the PHY security of buffer-aided FD decode-and-forward (DF) relaying. Unlike \cite{huang2015buffer, chen2015physical,shafie_FD_HDhybrid} and most existing work which assumed no direct link between Alice and Bob, we assume direct link communications and consider a FD AF relay node.

The contributions of this paper are summarized as follows
 \begin{itemize}
       \item We investigate the PHY security of the FD AF relay channel under direct link between Alice and both Bob and Eve.
       \item We show that the equivalent channels at the receivers due to transmissions from both Alice and the relay are intersymbol-interference (ISI) channels. Hence, the precoding and equalization processes complexities can be significantly simplified by using an orthogonal frequency division multiplexing (OFDM) modulator and demodulator at the transmitter and receiver side, respectively.
       \item We introduce the concept that the processing time (delay) at the relay node can be exploited to increase the delay spread of the equivalent Alice-Bob channel and Alice-Eve channel and change its characteristics and, hence, can be utilized to increase the system's PHY security by injecting artificial noise (AN).
       \item We derived the channel coefficients distributions across the frequency sub-channels and used them to derive the secrecy outage probability (SOP). 
\end{itemize}
\textit{Notation}: Unless otherwise stated, lower- and upper-case bold letters denote vectors and matrices, respectively. $\mathbf{I}_{\mathcal{B}}$ denotes the $\mathcal{B}\times \mathcal{B}$ identity matrix. $\mathbf{0}_{M \times N}$ denotes the  $M\times N$ all-zero matrix. $\mathbb{C}^{M \times N}$ and $\mathbb{B}^{M \times N}$ denote the set of all $M\times N$ complex and binary matrices, respectively. $\mathbb{C}$ denotes the set of all complex numbers. ${\rm diag}(\cdot)$ is the diagonal elements of the enclosed matrix. $(\cdot)^*/(\cdot)^\dagger$ denotes the Hermitian/transpose operation. $\mathbb{E}\{\cdot\}$ denotes expectation of the argument. $\lfloor\cdot \rfloor$ denotes the highest integer less than the argument. 

\section{Main Assumptions}
\begin{figure}
  \centering
    \includegraphics[width=0.8\columnwidth]{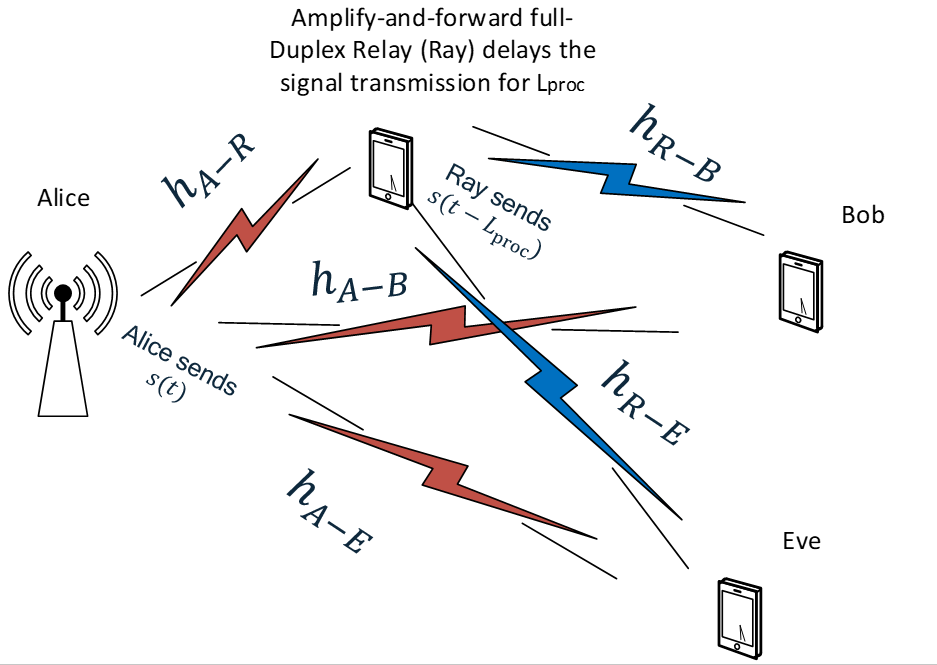}\\
  \caption{System model. The relay forwards the signal after imposing a processing delay of $L_{\rm proc}$.}\label{fig0}
  \vspace{-0.2cm}
  \end{figure}

As shown in Fig. \ref{fig0}, we consider a communication channel between a legitimate source node (Alice) and a legitimate destination node (Bob) in the presence of a single eavesdropping node (Eve) and an FD AF relay node (Ray). Relay nodes can be used in wireless or wired environments. In a wireless channel, FD relaying suffers from self interference which can reduce the achievable rates. We will focus on the wireless environment since it is the more general case where by setting some parameters, we can obtain the no self-interference wired channel scenario as a special case.

All nodes are equipped with a single antenna. Time is slotted into discrete equal-size time slots of duration $T$ seconds and the channel bandwidth is $W$ Hz. At the beginning of every time slot, Alice has a data packet to transmit. We assume that the direct link between Alice and Bob exists, unlike many exiting works including \cite{huang2015buffer,chen2015physical,shafie_FD_HDhybrid}. Hence, we assume that both Bob and Eve overhear both Alice's and Ray's transmissions. We consider the worst-case scenario where the eavesdropper's instantaneous channel state information (CSI) is unknown at the legitimate transmitting nodes.

\vspace{-0.5cm}
\subsection{Relay's Model}
\vspace{-0.05cm}
Ray is assumed to be an FD node where he can transmit and receive at the same time and over the same channel used by the source node. This relay can be DF or AF relay. Both relaying strategies are well-studied in literature and the main difference lies in the complexity versus data rate trade-offs. The complexity of AF relaying is much less than DF relaying, since the latter requires a full decoder at the Relay's receiver. In the high signal-to-noise ratio (SNR) regime, both the AF and DF achieve almost the same data rates. In this paper, we will focus on AF relaying due to its lower complexity. 

Ray is assumed to impose a processing time delay of $L_{\rm proc}\ge 1$. The minimum processing delay at an FD relay is $1$ symbol duration. In our proposed scheme, we assume that Ray can hold the data for a longer time to change the wireless channel characteristics which helps in hurting the eavesdropping channels by adding a strong AN signal to Alice's data signal.

\vspace{-0.4cm}
\subsection{Wireless Channel Model}
We consider a quasi-static flat-fading channel model where the entire codeword experiences a single channel realization \cite{bloch2011physical}. 
Alice transmits her data with a target secrecy rate, $\mathcal{R}$ bits/sec/Hz. The data transmission is secure whenever the target secrecy rate is below the achievable instantaneous secrecy rate. We assume that the channel coherence time is equal to a single slot duration where the channel coefficient of a link remains constant during a single time slot duration (codeword transmission duration) and changes independently from one time slot to another \cite{Marzban_keys_2020}. 

Let the channel coefficient between nodes $m_1$ and $m_2$ be denoted by $h_{m_1-m_2}$, and have a power gain of $|h_{m_1-m_2}|^2$, where $m_1,m_2\in\{\rm A,R,B,E\}$ denote Alice, Ray, Bob, and Eve, respectively. $h_{m_1-m_2}$ is a zero-mean complex Gaussian random variable with variance $\sigma_{m_1-m_2}^2$ and is assumed to be independent and identically distributed (i.i.d.) from one coherence time duration (time slot) to another. Each communication link is assumed to be corrupted by an additive white Gaussian noise (AWGN) process. Denote $n_{m_2}(t)$ as the AWGN process at receiver $m_2$ at symbol time $t$ with zero mean and power spectral density (PSD) $\kappa_{m_2}$ Watts/Hz resulting from the receiver thermal noise. In each time slot, the receivers are assumed to know their CSI with the transmitters which can be obtained through channel estimation and feedback. 
We assume that the average total transmit PSD at node $m_1$ per slot is $\mathcal{P}_{m_1}$ Watts/Hz. Hence, since the number of channel uses per coherence time is $\mathcal{B}_T=\lfloor WT\rfloor$ symbols, the average PSD per symbol is $\overline{\mathcal{P}_{m_1}}=\mathcal{P}_{m_1}/\mathcal{B}_T$ Watts/Hz. 
\vspace{-0.4cm}
\subsection{Received Signals}
We analyze the received signals at various nodes in the system. The received signal at Ray during time symbol $t\in\{0,1,2,\dots\}$ due to a transmission from Alice is given by
\begin{equation} \small \small
\begin{split} \small
\label{relay_sig}
y_{R}(t)&=h_{\rm A-R} s(t) +h_{\rm R-R} \mathcal{G}y_{\rm R}(t-L_{\rm proc})+n_{\rm R}(t)\\ &\!=\!h_{\rm A-R} s(t)  \!+\!\sum_{r\!=\!1}^{\lfloor\frac{\mathcal{B}_T}{L_{\rm proc}}\rfloor} \bigg[\mathcal{G}^r h_{\rm R-R}^r h_{\rm A-R} s(t-r L_{\rm proc}) \!\\& \,\,\,\,\,\,\,\,\,\,\,\,\,\,\,\,\,\,\,\,\,\,\,\,\,\,\,\,\,\,\,\,\,\,\,\,\,\,\,\,\,\,\,\,\,\,\,\,\,\,\,\ +\!\mathcal{G}^r h^r_{\rm R-R} n_{\rm R}(t-r L_{\rm proc}) \bigg]\!+\!n_{\rm R}(t)
\end{split}
\end{equation}
where $s(t)\in \mathbb{C}$, with $s(t)=0 \ \forall t< 0$, is the transmitted signal from Alice at symbol time $t$, $n_{\rm R}(t)$ is the AWGN signal at Ray, and $h_{\rm R-R}(t)\in \mathbb{C}$ is the residual self-interference at Ray after interference cancellation. The term $|h_{\rm R-R}(t)|^2 \overline{\mathcal{P}_{\rm R}}$ represents the symbol self-interference PSD.

Ray normalizes the received signal power and amplifies it by a weighting factor $\mathcal{G}$ and then forwards $\mathcal{G} y_{R}(t)$ to Bob with $\mathbb{E}\{\mathcal{G} y_{R}(t)\}=\overline{\mathcal{P}_{\rm R}}$. Hence, $\mathcal{G}$ is given by
\begin{equation} \small
\mathcal{G}=\sqrt{\frac{\overline{\mathcal{P}_{\rm R}}}{|h_{\rm A-R}|^2 \overline{\mathcal{P}_{\rm A}}  + |h_{\rm R-R}|^2 \overline{\mathcal{P}_{\rm R}} +\kappa_{\rm R}}}
\end{equation}

 

Since $h_{\rm R-R}$ is the residual self-interference, after filtering and interference cancellation, it has a very low variance, i.e., $\sigma^2_{\rm R-R}$ is very low. Hence, we can approximate the received signal and simply keep the first order term where the $r=1$ term is the dominant one across all terms for $r>1$. Hence,
\begin{equation} \small \small
\begin{split} \small
\label{relay_sig2x}
y_{R}(t)&=h_{\rm A-R} s(t) \\ & + \mathcal{G}\bigg[h_{\rm R-R} h_{\rm A-R} s(t-L_{\rm proc}) +h_{\rm R-R} n_{\rm R}(t-L_{\rm proc}) \bigg]+n_{\rm R}(t)
\end{split}
\end{equation}
The equivalent receive channel at Ray is a multi-tap channel with $2$ dominant (non-zero) taps at $0$ and $L_{\rm proc}$. Hence, the delay spread is $L_{\rm proc}$. Accordingly, the cyclic-prefix (CP) size has to be at least $L_{\rm proc}$ to remove the inter-block interference at Ray's receiver and flush the memory across data blocks.

Using \eqref{relay_sig2x}, the received signal at Bob/Eve is
\begin{equation} \small \small
\begin{split} \small
\label{Bob_sig}
y_{m}(t)&=h_{{\rm A}-m} s(t) +h_{R-m} \mathcal{G} y_{R}(t-L_{\rm proc})\\ &
= h_{{\rm A}-m} s(t)  +h_{R-m} \mathcal{G} \Bigg(h_{\rm A-R} s(t-L_{\rm proc}) \\ & + \mathcal{G}\bigg[h_{\rm R-R} h_{\rm A-R} s(t-2 L_{\rm proc}) +h_{\rm R-R} n_{\rm R}(t-2 L_{\rm proc})\bigg] \\ & +n_{\rm R}(t-L_{\rm proc})\Bigg) + n_{m}(t)
\end{split}
\end{equation}
where ${m}\in \{\rm B,E\}$, and $n_{\rm m}(t)$ is the AWGN at node ${m}$ receiver (i.e., Bob/Eve receiver) at time $t$ with zero mean and PSD of  $\kappa_{\rm m}$. The equivalent receive channel at Bob (similarly at Eve) is a multi-tap channel with $3$ dominant taps at $0, L_{\rm proc}$, and $2L_{\rm proc}$. However, the tap at $2L_{\rm proc}$ is multiplied by $\mathcal{G} h_{\rm R-B} h_{\rm R-R} h_{\rm A-R}$ with $h_{\rm R-R}$ having very low variance. Hence, effectively, the delay spread is $L_{\rm proc}$. 

Rearranging the expression in Eqn. \eqref{Bob_sig}, the vector representation of the signal at Bob/Eve is 
\begin{equation} \small \small
\begin{split} \small
\label{Bob_sigVec}
\mathbf{y}_{m}&= \mathbf{H}_{{\rm AR}-m}\mathbf{s} + \tilde{\mathbf{n}}_{{\rm R}-m} + \mathbf{n}_{\rm m}
\end{split}
\end{equation}
where $\mathbf{s}=[s(0),s(1),\dots,s(\mathcal{B}_T-1)]^\dagger \in \mathbb{C}^{\mathcal{B}_T\times 1}$ is the data symbols vector, $\mathbf{n}_{m} \in \mathbb{C}^{\mathcal{B}_T\times 1}$ is receiver $m$ AWGN vector, and $\mathbf{H}_{{\rm AR}-m}$ is the Toeplitz channel matrix  between the transmitters (Alice and Ray) and Bob/Eve. The vector $\tilde{\mathbf{n}}_{{\rm R}-m} \in \mathbb{C}^{\mathcal{B}_T\times 1}$ has the following form. Elements from $1$ to $L_{\rm proc}$ are zeros. Moreover, elements from $L_{\rm proc}+1$ to  $2L_{\rm proc}$ are $h_{{\rm R}-m} \mathcal{G}  n_{\rm R}(0),h_{{\rm R}-m} \mathcal{G}  n_{\rm R}(1),\cdots,h_{{\rm R}-m} \mathcal{G}  n_{\rm R}(L_{\rm proc}-1)$. Finally, elements from $2L_{\rm proc}+1$ to $\mathcal{B}_T$ are $h_{{\rm R}-m} \mathcal{G}^2 h_{\rm R-R} n_{\rm R}(0)+h_{{\rm R}-m} \mathcal{G} n_{\rm R}(L_{\rm proc}), h_{{\rm R}-m} \mathcal{G}^2 h_{\rm R-R} n_{\rm R}(1)+h_{{\rm R}-m} \mathcal{G} n_{\rm R}(L_{\rm proc}+1),\cdots,h_{{\rm R}-m} \mathcal{G}^2 h_{\rm R-R} n_{\rm R}(\mathcal{B}_T-2L_{\rm proc}-1)+h_{{\rm R}-m} \mathcal{G}n_{R}(\mathcal{B}_T-L_{\rm proc}-1)$. 

\begin{figure}
   \vspace{-0.8cm}
  \centering
    \includegraphics[width=0.8\columnwidth]{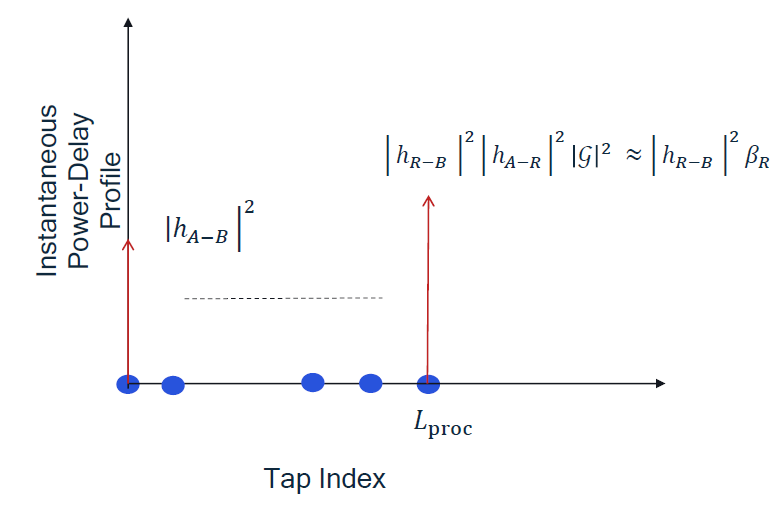}\\
  \caption{Channel power-delay profile realization of the equivalent channel at Bob. In the figure, $\beta_{\rm R}$ indicates the power ratio between Ray's PSD and Alice's PSD. The first tap is from Alice's transmission while the second one is from Ray's transmission.}\label{fig1}
  \vspace{-0.4cm}
  \end{figure}

Assuming that Ray performs full self-interference cancellation, i.e., $\overline{\mathcal{P}_R} \mathbb{E}\{|h_{\rm R-R}|^2\}$ is negligible, Ray's received signal is given by
\begin{equation} \small \small
\begin{split} \small
\label{relay_sig2}
y_{R}(t)&=h_{\rm A-R} s(t) +n_{\rm R}(t)
\end{split}
\end{equation}
Hence, the received signal at Bob/Eve is then given by
\begin{equation} \small \small
\begin{split} \small
\label{Bob_sig_sim}
y_{m}(t)
& = h_{{\rm A}-m} s(t)   +h_{R-m} \mathcal{G} h_{\rm A-R} s(t-L_{\rm proc})  \\ & +h_{R-m}\mathcal{G} n_{\rm R}(t-L_{\rm proc})+ n_{m}(t)
\end{split}
\end{equation}
where ${m}\in \{\rm B,E\}$ and $\mathcal{G}=\sqrt{\frac{\overline{\mathcal{P}_{\rm R}}}{|h_{\rm A-R}|^2 \overline{\mathcal{P}_{\rm A}} +\kappa_{\rm R}}}$. The equivalent channel due to the transmission of Alice and Ray signals to Bob becomes simply a two-tap channel with a delay spread of  $L_{\rm proc}$; all other taps are zeros, see Fig. \ref{fig1}.

\vspace{-0.1cm}
\section{Links' Rates and Secrecy Rate}
\vspace{-0.05cm}
To avoid inter-block interference between one block of data (codeword) and the other blocks across different coherence times, Alice has to pad each data block with a guard. If this guard is the last $N_{\rm cp}$ symbols of the data block, the equivalent channel matrix between the transmitters (Alice and Ray) and Bob becomes \emph{circulant} and is given by
\begin{equation} \small \small
\begin{split} \small
\tilde{\mathbf{H}}_{\rm AR-B}=\mathbf{R}^{\rm cp}\mathbf{H}_{\rm AR-B}\mathbf{T}^{\rm cp}
\end{split}
\end{equation}
where $\mathbf{T}^{\rm cp} \in \mathbb{B}^{(\mathcal{B}+N_{\rm cp})\times \mathcal{B}}$, $\mathbf{R}^{\rm cp} \in \mathbb{B}^{\mathcal{B} \times (\mathcal{B}+N_{\rm cp})}$ are the CP insertion and removal matrices, respectively, $\mathcal{B}$ is the data block size, and $\mathbf{H}_{\rm AR-B} \in \mathbb{C}^{(\mathcal{B}+N_{\rm cp}) \times \mathcal{B}}$ is a Toeplitz matrix whose first column is $[h^*_{\rm A-B},\underbrace{0,\cdots,0}_{L_{\rm proc}-1},\sqrt{\beta_{\rm R}}h^*_{\rm R-B},0,\dots,0]^*$ with $\beta_{\rm R}$ indicating the power ratio between Ray's PSD and Alice's PSD.

Assuming complex Gaussian data symbols, the Bob's receiver achieved rate in bits/sec/Hz is 
\begin{equation} \small \small
\begin{split} \small
\label{cocowawa2v233ccc}
\mathcal{R}_{\rm B}=\frac{1}{\mathcal{B}+N_{\rm cp}}\log_2 \det\left(\mathbf{I}_{\mathcal{B}}+ \frac{\overline{\mathcal{P}_A}}{\kappa_{\rm B}}\tilde{\mathbf{H}}_{\rm AR-B}^* \tilde{\mathbf{H}}_{\rm AR-B}\right)
\end{split}
\end{equation}
where $N_{\rm cp}\ge L_{\rm proc}$ and $\mathcal{B}+N_{\rm cp}$ is the number of channel uses within a coherence time. Since $\tilde{\mathbf{H}}_{\rm AR-B}$ is a circulant matrix, the optimal data encoding and decoding is the discrete-Fourier Transform (DFT) (i.e., nodes use OFDM modulator/demodulator). That is, $\tilde{\mathbf{H}}_{\rm AR-B}$ can be decomposed as $\tilde{\mathbf{H}}_{\rm AR-B}=\mathbf{F}^*\mathbf{\Lambda} \mathbf{F}$, where $\mathbf{F}$ is the DFT matrix, 
while the elements of the eignevalues matrix, $\mathbf{\Lambda}$, are the DFT coefficients of the channel taps. On the other hand, the Eve's receiver achieved rate can be expressed as 
\begin{equation} \small \small
\begin{split} \small
\label{Eve_data_Rate_noAN}
\mathcal{R}_{\rm E}=\frac{1}{\mathcal{B}+N_{\rm cp}}\log_2 \det\left(\mathbf{I}_{\mathcal{B}}+ \frac{\overline{\mathcal{P}_A}}{\kappa_{\rm E}}\mathbf{H}_{\rm AR-E}^* \mathbf{H}_{\rm AR-E}\right)
\end{split}
\end{equation}
where $\mathbf{H}_{\rm AR-E} \in \mathbb{C}^{(\mathcal{B}+N_{\rm cp}) \times (\mathcal{B}+N_{\rm cp})}$ is a Toeplitz matrix whose first column is $[h^*_{\rm A-E},\underbrace{0,\cdots,0}_{L_{\rm proc}-1},\sqrt{\beta_{\rm R}}h^*_{\rm R-E},0,\dots,0]^*$. 

The system's secrecy rate and secure throughput are given, respectively, by
\begin{equation} \small \small
\begin{split} \small
\label{total_Rate}
\mathcal{R}^{\rm sec}_{\rm A-B}&=[\mathcal{R}_{\rm A-B}-\mathcal{R}_E]^+
\\
\mathcal{T}^{\rm sec}_{\rm A-B}&=\mathcal{R} \Pr\{\mathcal{R}^{\rm sec}_{\rm A-B}\ge \mathcal{R}\}=\mathcal{R} (1-P^{\rm outage}_{\rm sec})
\end{split}
\end{equation}
and $P^{\rm outage}_{\rm sec}=\Pr\{\mathcal{R}^{\rm sec}_{\rm A-B}< \mathcal{R}\}$ is the SOP, which is monotonically non-increasing with $\mathcal{R}$.

From the rate expressions at both Bob and Eve receivers, we have the following observations.
\begin{itemize}
\item To decode the entire $\mathcal{B}$ codeword of data within a coherence time duration, the receivers need to jointly decode all symbols.
\item To reduce the complexity and since ISI can be mitigated by implementing an OFDM modulator/demodulator at transmitter/receiver side, we can convert the frequency-selective channel into a group of flat-fading sub-channels/sub-carriers.
\item By adding CP guard at Alice and removing it at Bob, Bob can remove the interference from one OFDM symbol to another.
\item By increasing the processing delay and adding a CP of size at least equal to the channel delay spread ($N_{\rm cp}$ is equal to $L_{\rm proc}$), we increase the null space dimension of the equivalent Alice-Bob channel matrix. This interesting design can be used to  introduce an AN signal at Alice and its impact will be removed at Bob after removing CP. More specifically, the AN does not degrade Bob's receiver by selecting the AN precoder matrix based on the following condition:
\begin{equation} \small \small
\begin{split} \small
&\mathbf{R}^{\rm cp}\mathbf{H}_{\rm AR-B}\mathbf{U}=\mathbf{0}_{\mathcal{B}_T \times N_{\rm cp}} \rightarrow \mathbf{U}={\rm null}(\mathbf{R}^{\rm cp}\mathbf{H}_{\rm AR-B})
\end{split}
\end{equation}
where $\mathbf{R}^{\rm cp}\mathbf{H}_{\rm AR-B}$ is a wide matrix (i.e., number of columns is larger than number of rows) with size $\mathcal{B}\times (\mathcal{B}+N_{\rm cp})$. The matrix $\mathbf{U}$ has a null space with dimension equal to $N_{\rm cp}$ with only $L_{\rm proc}$ useful AN streams that can hurt Eve \cite{8615994}. Hence, the interference covariance matrix rank at Eve is $L_{\rm proc}$.
\item Ray's processing delay is controlling the channel memory. Effectively, the channel impulse responses (CIRs) will have only $2$ taps but the delay spread is still controlled by Ray.
\item By optimizing the processing delay at Ray, Alice can enhance her secure transmissions, but there is a trade-off in such selection where if the processing time is very large, undesirable data rate losses are introduced  due to the CP overhead and power penalty. Hence, there is an optimum value for the processing time, $L_{\rm proc}$, that maximizes the system's secure throughput.
\end{itemize}

By adding a CP/guard interval and implementing the AN injection scheme, the achievable rate at Eve's receiver is
\begin{equation} \small \small
\begin{split} \small
\label{Eve_data_Rate_AN}
\mathcal{R}_{\rm E}&=\frac{1}{\mathcal{B}+N_{\rm cp}}\log_2 \det\Bigg(\mathbf{I}_{\mathcal{B}}+ \frac{\mathcal{P}^{\rm data}_{\rm A}}{\mathcal{B}+N_{\rm cp}}\tilde{\mathbf{H}}_{\rm AR-E} \tilde{\mathbf{H}}_{\rm AR-E}^* \\ & \,\,\,\,\,\,\,\,\,\ \times \left(\kappa_{\rm E} \mathbf{I}_{\mathcal{B}} +\frac{\mathcal{P}^{\rm AN}_{\rm A}}{L_{\rm proc}} \mathbf{R}^{\rm cp}\mathbf{H}_{\rm AR-E}\mathbf{U}\left(\mathbf{R}^{\rm cp}\mathbf{H}_{\rm AR-E}\mathbf{U}\right)^* \right)^{-1}\Bigg)
\end{split}
\end{equation}
where $\mathcal{P}^{\rm AN}_{\rm A}/L_{\rm proc}$ is the AN symbol power with $\mathcal{P}^{\rm AN}_{\rm A}=\theta \mathcal{P}_{\rm A}$ and $0\le \theta\le 1$. Furthermore, the total data power becomes $\mathcal{P}^{\rm data}_{\rm A}=\mathcal{P}_{\rm A}-\mathcal{P}^{\rm AN}_{\rm A}=(1-\theta) \mathcal{P}_{\rm A}$. It was shown in many works that, at moderate-to-high SNR regimes, the optimal value allocation $\theta$ is $\theta=1/2$ \cite{7543526}.

We note that the AN complicate the receiver at Eve since it couples the sub-channels, i.e., it creates correlation between the data across sub-channels since the AN is generated by a linear precoder. Hence, the best decoder at Eve is the joint detection of all data across all sub-channels. However, Eve might decide to just use the conventional OFDM detection where a single-tap equalization per sub-channel is used, and this will hurt her achievable rate more \cite{8615994,7543526}.

Since the block size, $\mathcal{B}$, is very large, to reduce decoding complexity, Alice and Bob can split the data into small blocks then add CP/guard interval to flush the memory and enable Bob to decode each data block individually. Alice divides the data into blocks of $N$ complex symbols and each block is cyclically extended by padding its beginning with a repetition of its last $N_{\rm cp}$ symbols.\footnote{Since the effective ISI channel created by Alice and Ray transmissions is always a two-tap channel, a Viterbi receiver would allow a symbol-by-symbol detection for discrete input data instead of block processing as in OFDM.} Within a coherence time, Alice sends $\lfloor\mathcal{B}_T/(N+N_{\rm cp})\rfloor$ block of symbols. In the derived achievable rate expressions, we replace $\mathcal{B}$ with~$N$.

\section{OFDM Modulation Scenario and Proposed Optimization}
In OFDM case, the $N$ data symbols are transmitted from Alice to Bob. Each OFDM symbol is padded with a CP of size $N_{\rm cp}$. Assume an OFDM modulator at Alice and OFDM demodulator at Bob\footnote{Ray would need a modulator and a demodulator if DF relaying is utilized.}, the channel matrix between the transmitters (Alice and Ray) and the receiving node $m\in\{\rm B,E\}$ becomes a diagonal matrix as follows
\begin{equation} \small
    \begin{split}
        \mathbf{H}^{\rm freq}_{{\rm AR}-m}=\mathbf{F}\mathbf{R}^{\rm cp} \mathbf{H}_{{\rm AR}-m} \mathbf{T}^{\rm cp} \mathbf{F}^*
    \end{split}
\end{equation}
where $\mathbf{h}_{{\rm AR}-m} \in \mathbb{C}^{(L_{\rm proc} +1) \times 1}$ is the CIR of the link between the transmitters (Alice and Ray) and node $m \in \{\rm B,E\}$. Note that $N$ is the FFT size and represents the number of sub-channels and $\mathcal{B}_T=N+N_{\rm cp}$ represents the OFDM symbol duration where $N_{\rm cp} \ge L_{\rm proc}$ is the CP size. The channel vector across all sub-channels is given~by
\begin{equation} \small
    \begin{split}
        \mathbf{h}^{\rm freq}_{{\rm AR}-m}&={\rm diag}(\mathbf{H}^{\rm freq}_{{\rm AR}-m})=\sqrt{N}\mathbf{F}
\begin{bmatrix}
\mathbf{h}_{{\rm AR}-m}\\
\mathbf{0}_{N-(L_{\rm proc}+1)}
\end{bmatrix}\\& =\mathbf{f}_0
h^0_{{\rm A}-m}+\mathbf{f}_{L_{\rm proc}}
h^{L_{\rm proc}}_{{\rm R}-m}
    \end{split}
\end{equation}
where $\mathbf{f}_i$ is the $i$-th column of the DFT/FFT matrix $\sqrt{N} \mathbf{F}$. The correlation matrix among the sub-channels is given by $\sigma_{{\rm A}-m}^2 \mathbf{f}_0\mathbf{f}_0^*+\sigma_{{\rm R}-m}^2 \beta_{\rm R}\mathbf{f}_{L_{\rm proc}}\mathbf{f}_{L_{\rm proc}}^*$. When $\mathcal{P}_{\rm R}=\mathcal{P}_{\rm A}$ (i.e., $\beta_{\rm R}=1$) and the channel coefficients have unit variances, the channels experience uniform power-delay profiles with $\mathbf{F}_1 \mathbf{F}^*_1=\mathbf{f}_0\mathbf{f}_0^*+\mathbf{f}_{L_{\rm proc}}\mathbf{f}_{L_{\rm proc}}^*$

Denoting $h^{\ell}_{{\rm AR}-m}$ as the $\ell$-th tap of the equivalent CIR between transmitters ${\rm AR}$ and receiving node $m \in \{\rm B,E\}$, the channel coefficient of sub-channel $k$ is given by
\begin{equation} \small
    \begin{split}
H^k_{{\rm AR}-m}&=\sum_{\ell=0}^{N-1} h^{\ell}_{{\rm AR}-m} \exp\left(-j 2\pi \frac{\ell k}{N}\right)\\ &=h^{0}_{{\rm A}-m}+h^{L_{\rm proc}}_{{\rm R}-m} \exp\left(-j 2\pi \frac{L_{\rm proc} k}{N}\right)
    \end{split}
\end{equation}
 Hence, the distribution of each sub-channel is complex Gaussian random variable with zero mean and variance $\sum_{\ell=0}^{\nu} \sigma^2_{k,\ell} $ where $\sigma^2_{k,\ell}$ is the variance of the $\ell$-th CIR tap and $\nu$ is the number of non-zero taps. In our case, $\nu=2$ taps. The variance of $H^k_{{\rm AR}-m}$ is $\sigma_{{\rm A}-m}^2  +\sigma_{{\rm R}-m}^2 \beta_{\rm R}$ for node $m \in \{\rm B,E\}$.

  Using the SOP expression in \cite[Eqn. (7)]{barros2006secrecy} for a single-tap channel as the per sub-channel SOP, and substituting with our system's parameters, the SOP per sub-channel $k$ is given~by
 \begin{equation} \small
 \begin{split}
 \label{pklolkk}
P^{\rm outage}_{{\rm sec},k}\!
\!=\!1&-\frac{\sigma_{\rm A-B}^2 \frac{1}{\kappa_{\rm B}}+\sigma_{\rm R-B}^2\frac{\beta_{\rm R}}{\kappa_{\rm B}}}{\sigma_{\rm A-B}^2 \frac{1}{\kappa_{\rm B}}+\sigma_{\rm R-B}^2\frac{\beta_{\rm R}}{\kappa_{\rm B}}+2^{\mathcal{R}}\frac{\sigma_{\rm A-E}^2+ \sigma_{\rm R-E}^2\beta_{\rm R}}{\kappa_{\rm E}+\delta_k}}\\ & \times \exp\left(-\frac{2^\mathcal{R}-1}{ \gamma_{{\rm B}}}\right)
\end{split}
\end{equation}
where $\gamma_{\rm B}= \frac{\mathcal{P}^{\rm data}_{\rm A}(\sigma_{\rm A-B}^2 +\sigma_{\rm R-B}^2\beta_{\rm R})}{\kappa_{\rm B}}$ and $\delta_k$ is the AN interference power at sub-channel $k$ given by the $(k,k)$ element of $\frac{\mathcal{P}_{\rm A}^{\rm AN}}{L_{\rm proc}}\mathbf{R}^{\rm cp}\mathbf{H}_{\rm AR-E}\mathbf{U}\left(\mathbf{R}^{\rm cp}\mathbf{H}_{\rm AR-E}\mathbf{U}\right)^*$. Eqn. \eqref{pklolkk} reduces to
 \begin{equation} \small
 \begin{split}
 \label{pklolkkx}
P^{\rm outage}_{{\rm sec},k}\!
&\!=\!1-\alpha_{k} \exp\left(-\frac{2^\mathcal{R}-1}{ \gamma_{{\rm B}}}\right)
\end{split}
\end{equation}
where
\begin{equation} \small
 \begin{split}
 \label{pklolkkxx}
\alpha_k=\frac{\sigma_{\rm A-B}^2 \frac{1}{\kappa_{\rm B}}+\sigma_{\rm R-B}^2\frac{\beta_{\rm R}}{\kappa_{\rm B}}}{\sigma_{\rm A-B}^2 \frac{1}{\kappa_{\rm B}}+\sigma_{\rm R-B}^2\frac{\beta_{\rm R}}{\kappa_{\rm B}}+2^{\mathcal{R}}\frac{\sigma_{\rm A-E}^2+ \sigma_{\rm R-E}^2\beta_{\rm R}}{\kappa_{\rm E}+\delta_k}}
\end{split}
\end{equation}
The term $\exp\left(-\frac{2^\mathcal{R}-1}{ \gamma_{{\rm B}}}\right)$ is the probability that the there is no outage at sub-channel $k$ when there is no eavesdropping. The parameter $\alpha_{k}$ can be seen as the increase of the outage events due to the presence of eavesdropping.

As Alice and Ray transmit PSD levels increase, $\exp\left(-\frac{2^\mathcal{R}-1}{ \gamma_{{\rm B}}}\right)$ becomes closer to $1$, i.e., the outage probability due to fading is close to zero. However, due to the security constraint and presence of Eve, the term $\alpha_k$ is independent of Alice and Ray transmit power. Hence, eventually, the SOP saturates at $P^{\rm outage}_{{\rm sec},k}=\alpha_k$. If $\delta_k=0$, i.e., no AN injection, this number will remain high. Hence, an efficient way to increase $\alpha_k$, as we did here, is by introducing an AN signal and by utilizing Ray to increase the processing delay which, in turn, increases the interference covariance matrix rank at Eve and significantly degrades her rate. 

Our aim is to maximize the secure throughput, which requires the eavesdropping links' statistics only, through the optimization of Ray's processing time $1\le L_{\rm proc} \le N$. The optimization problem is stated as follows:
\begin{equation} \small \small
\begin{split} \small
\label{e2e22}
\underset{1\le L_{\rm proc} \le N}{\max:} \ \mathcal{T}^{\rm sec}_{\rm A-B}=\mathcal{R} (1-P^{\rm outage}_{\rm sec})
\end{split}
\end{equation}
The optimization problem is a function of one integer parameter, $L_{\rm proc}$. The optimal solution of $L_{\rm proc}$ does not change as long as the system’s average parameters remain unchanged.

   \begin{figure}
   \vspace{-1cm}
  \centering
    \includegraphics[width=0.8\columnwidth]{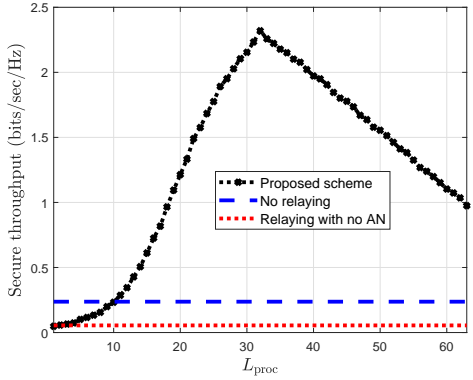}\\
  \caption{Secure throughput versus $L_{\rm proc}=N_{\rm cp}$.}\label{Nsecrecy outage probabilityVsNcp1}
   \vspace{-0.2cm}
  \end{figure}
  \vspace{-0.3cm}
\section{Numerical Results and Simulations}
We provide some numerical results to evaluate the proposed ideas and show the secure throughput gains as we increase Ray's processing time. The fading channels are assumed to be complex circularly-symmetric Gaussian random variables with zero mean and unit variance. Unless stated otherwise, we consider an OFDM system with $N=64$, $N_{\rm cp} =16$, $\theta=1/2$, $\mathcal{R}=4$ bits/sec/Hz, and $\overline{\mathcal{P}_{m_1}}/\kappa_{m_2}= 30$ dB.

In Fig. \ref{Nsecrecy outage probabilityVsNcp1}, we show the gains of our proposed scheme relative to three benchmarks, namely, no processing time controlling case where $L_{\rm proc}$ is set to $1$, no relaying case, and relaying case with no AN injection. Fig. \ref{Nsecrecy outage probabilityVsNcp1} shows the gains of increasing the processing time at Ray in terms of secure throughput. As $L_{\rm proc}$ increases, the secure throughput increases until a peak is reached then the secure throughput saturates. This is intuitive since increasing $L_{\rm proc}$ increases the number of AN streams. In other words, the rank of the AN interference covariance matrix at Eve increases to $L_{\rm proc}$. However, at some point, increasing $L_{\rm proc}$ reduces the prelog factor of the links' rates since $N_{\rm cp}=L_{\rm proc}$. In addition, we observe that the secure throughput for the benchmark of no relaying outperforms the relaying with no AN-injection case as the latter introduces ISI at Bob from Alice and Ray concurrent transmissions.

\begin{figure}
   \vspace{-1cm}
  \centering
    \includegraphics[width=0.8\columnwidth]{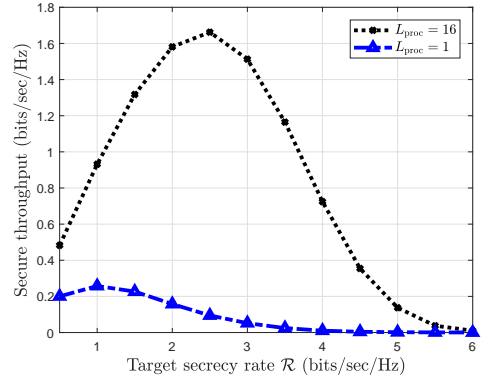}\\
  \caption{Secure throughput versus $\mathcal{R}$ at different processing times}\label{VersusTargetRate}
   \vspace{-0.2cm}
  \end{figure}

Fig. \ref{VersusTargetRate} shows that the secure throughput increases with the target secrecy rate $\mathcal{R}$ until the secure throughput reaches a peak then decreases again. This is due to the fact that the secure throughput is equal to $\mathcal{R}$ (increasing function of $\mathcal{R}$) multiplied by $1-P^{\rm outage}_{\rm sec}$ (decreasing function with $\mathcal{R}$). Fig. \ref{VersusTargetRate} is plotted for two values of $L_{\rm proc}$ to further emphasize the gain of Ray's processing time increase. At $L_{\rm proc}=16$, the highest secure throughput is almost $1.66$ bits/sec/Hz, while under $L_{\rm proc}=1$ (relay does not add further delays), the highest secure throughput is $0.25$ bits/sec/Hz. This shows almost $564\%$ gain under increased processing time at Ray because of the increased covariance matrix rank of the induced interference due to the injected AN. 


  \vspace{-0.2cm}
\section{Conclusions}
We proposed a new method to secure the FD AF relaying channel. We showed the impact of the processing time at the relay on improving the PHY security. We showed that the complexities of the modulation and demodulation processes at the various nodes can be significantly reduced by using OFDM, even when individual links are experiencing single-tap channels due to the induced ISI from FD AF relaying. By injecting artificial noise and controlling the processing delay, we can enhance the secure throughput significantly. For example, under the proposed scheme, at $L_{\rm proc}=16$, the highest secure throughput is almost $1.66$ bits/sec/Hz, while under $L_{\rm proc}=1$ (relay does not add further delays), the highest secure throughput is $0.25$ bits/sec/Hz. This shows a significant gain, almost $564\%$ due to increased relay's processing time because of the increased covariance matrix rank of the induced interference due to the injected AN.
\vspace{-0.3cm}
\bibliographystyle{IEEEtran}
\bibliography{IEEEabrv,term_bib}
\end{document}